\documentclass[twocolumn,showpacs,showkeys,pre,floatfix,amsmath]{revtex4}
\usepackage{graphicx}
\newcommand{\VEC}[1]{\overrightarrow{#1}}

\begin{document}
\newcommand{\singlefiguresize}{0.95\columnwidth}



\title{Slow relaxation to equipartition in spring-chain systems}


\author{Tetsuro Konishi$^1$, Tatsuo Yanagita$^2$}
\affiliation{$^1$ Department of Physics, Nagoya University, Nagoya 464-8602, Japan,}

\affiliation{$^2$ Research Institute  for Electronic Science, Hokkaido University, Sapporo 001-0020, Japan}


\date{\today}

\begin{abstract}
In this study, 
one-dimensional systems of masses connected by springs,
i.e., spring-chain systems,  are investigated numerically.
The average kinetic energy of chain-end particles of these systems
is larger than that of other particles, 
which is similar to the behavior observed for systems made of masses connected by rigid links.
The energetic motion of the end particles is, however, transient, 
and the system relaxes to thermal equilibrium after a while, where
the average kinetic energy of each particle is the same, that is, equipartition
of energy is achieved.
This is in contrast to the case of systems made of masses connected by rigid links,
where  the energetic motion of the end particles is observed in equilibrium.
The timescale of relaxation  estimated by simulation increases rapidly 
with increasing spring constant.
The timescale is also estimated using the Boltzmann-Jeans theory and 
is found to be in quite good agreement with that obtained by the simulation.
\end{abstract}


\maketitle

\section{Introduction}
Constrained systems are those having constraints on their degrees of freedom.
When a constraint is imposed on  spatial coordinates, the constraint is called a
``holonomic constraint''\cite{goldstein}. 
Constrained systems are
useful and simple, because of which they are 
 widely used as model systems.
An example of such a model system is a 
freely jointed chain~\cite{Kramers-chain,chain-letter-JSTAT-2009,Mazers-chain-pre-1996},
which 
is a model composed of one-dimensional chain (1D) of $N$ masses
such that  
the distances between adjacent masses are constant. 
The freely jointed chain is known  as a simplified model of polymers.
In computational software packages 
such as
CHARMM~\cite{CHARMM} and AMBER~\cite{AMBER}
for molecular dynamics calculations, 
the SHAKE and RATTLE algorithms enable one to treat model systems as constrained ones 
by setting distances  between atoms as constant. 
As a result,
the computational task can be made much easier, and then, 
physically important results can be obtained.
In molecular dynamics calculations, water molecules are often treated as 
having a 
fixed shape, where the length of bonds between hydrogen and oxygen atoms
is fixed~\cite{TIP4P}.
In both cases, we replace the bond between atoms 
with a rigid link when the frequency of bond vibration is extremely high.

It is known that
in a constrained system, the equipartition of energy occurs
in a somewhat complicated way, and the average kinetic energies
of particles $\langle 1/2 m_i v_i^2 \rangle$ can take different values for 
particles located at different places in the system~\cite{ryckaert-1986}.
In the generalized form of 
equipartition~\cite{Tolman-book,kubo-book}, 
what is equal among degrees of freedom is not $\langle 1/2 m_i v_i^2 \rangle$ but
$\langle 1/2 p_i \frac{\partial K}{\partial p_i} \rangle$, 
where $p_i$ is the momentum conjugate to the generalized coordinate $q_i$ 
of the $i$'th degree of freedom, and $K$ is the kinetic energy of the system.
In constrained systems, $K$ depends on coordinates and $p_i$ is no longer 
equal to $m_i v_i$; therefore $ 1/2 p_i \frac{\partial K}{\partial p_i} $, 
whose average takes the same value for all $i$, is not equal to $1/2 m_i v_i^2$.

Recently, we  found that for a chain-type system, termed a  planar chain model, 
the average kinetic energy of each particle differs systematically; 
that is,  particles near  both the ends
of the chain have relatively large average kinetic
energies~\cite{chain-letter-JSTAT-2009}.
This model consists of masses connected by rigid links;  since the 
 distances between adjacent masses are fixed in this model,
it is a constrained system.
It is the constraint that induces the nonuniformity of 
average kinetic energies.
The abovementioned energetic motion of end particles  observed in the model
would be useful in understanding the dynamical behavior of 
chain-type systems such as polymers~\cite{doi-edwards}, DNAs, 
proteins~\cite{leitner-straub-protein-2009}, and some artificial objects
such as manipulator arms of spacecraft.

Thus far, we have described  the behavior of constrained systems.
However, strictly speaking,  a constrained system
or rigid link does not exist in the real world. 
When the potentials of a system are somewhat steep, i.e., spring constants
are quite large and the frequency of bond vibration is reasonably high, we 
approximate
the bonds with rigid links.
A rigid link or  holonomic constraint is an idealized limit of a stiff spring.

However, if we replace the rigid links with springs, say spring-chain model,
the usual expression of 
equipartition of energy, i.e.,
 $\langle \frac{1}{2}m_i v_i^2 \rangle = \frac{D}{2}k_B T$, holds
regardless of the magnitude of the spring constant $k$,
where $D$ denotes the spatial dimension.
In other words, although a spring-chain system, made by
replacing rigid links in the planar chain system with springs,
appears to behave like 
the planar chain model when the spring constants are large, 
the behavior of energetic motion of end particles cannot be reproduced by
applying equilibrium statistical mechanics
to the model.

Then 
it would be interesting to know 
 whether  the large average kinetic energies of end particles
observed in the   planar chain model can also be found in the real world
or whether it is  an artifact observed only in mathematical models
and is never observed in the real world.
If the former is true, then we can expect to observe an interesting feature that
the energy distribution of many-body systems shows nonuniform behavior.
Since it is natural to consider that the dynamical behavior of a stiff spring 
is similar to that of a rigid link at least for a finite time,
the solution to  the problem will be the knowledge of  the relaxation properties to equilibrium.

With this background, the aim
of this study is
 to examine the property of relaxation to 
equipartition for the spring-chain system,
particularly for a large value of the spring constant $k$.
We measure the relaxation time $t_{relax}$ and investigate its relation 
with the spring constant $k$.
Further, 
we estimate $t_{relax}$ by the Boltzmann-Jeans theory
and compared it with the value measured.

This paper is organized as follows. In Sec.II we introduce the model,
the spring-chain system. In Sec.III we briefly describe 
the method of numerical computation. The results are shown in Sec.IV.
The final section is devoted to the summary and discussion.
\section{Model}
\begin{figure}
  \centering
  \includegraphics[width=0.95\columnwidth]{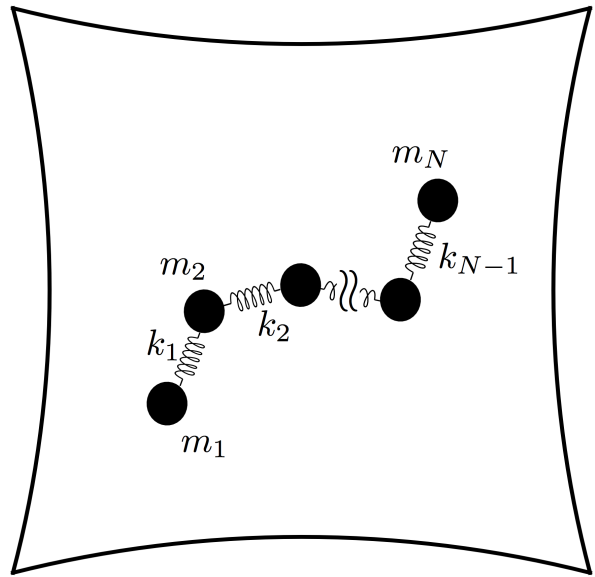}
  \caption{A schematic illustration of spring-chain model.}
  \label{fig:spring-chain}
\end{figure}

We now
introduce the spring-chain system.
It is composed of $N$ particles (masses) connected by
$N-1$ massless springs. The masses can rotate smoothly on the $xy$-plane, as shown in Fig.~\ref{fig:spring-chain}.
The system is defined by the following Lagrangian $L$:
\begin{eqnarray}
  L&=&\sum_{i=1}^N \frac{m_i}{2}\left(\dot x_i^2 + \dot y_i^2\right)
- \sum_{i=1}^{N-1}\frac{k_i}{2}\left\{\left|\VEC{r_{i+1}}-\VEC{r_i}\right| - \ell_i\right\}^2 \nonumber\\
 &-&  U(\left\{ \VEC{r_i} \right\} ) \ , 
\label{eq:spring-chain}
\end{eqnarray}
where 
$m_i$ is the mass of the $i$'th particle, 
$\VEC{r_i}\equiv(x_i,y_i)$ represents the position of  the $i$'th particle,  
$k_i$ and $\ell_i$ are the spring constant and  natural length of the $i$'th spring,
respectively.
$U$ is an external potential.
In this paper we consider the case of $m_i = m$, $k_i = k$,  and $\ell_i = \ell$ 
for all $i$. We set $m = 1$ and $\ell =1$.
The spring-chain model is a kind of beads-type models, which are 
used as models of polymer and protein~\cite{Rouse,doi-intro-polymer,wales-landscape-book,wolynes-science-1995,thirumalai-BLN-pnas-1990}. 
\section{Method of Numerical Integration}
We integrate the equation of motion of the model by a 
fourth-order symplectic integrator that is the composition of three successive
second-order symplectic integrators. 
External potential~\cite{chain-letter-JSTAT-2009}
$U(\VEC{r})=0.01 \sum_{j=1}^{N_{wall}}
\left|\left|\VEC{r} - \VEC{R}_j\right|-a\right|^{-6}$  
is applied in order to break the rotational symmetry and thus 
prevent
the conservation of angular momentum.
Here
$N_{wall}=4$, $a=4N\ell$, 
$\vec{R_j}\equiv (R,0)$, $ (0,R)$, $ (-R,0)$, $(0,-R)$,
$R\equiv N\ell + \sqrt{a^2-N^2\ell^2}$.

Throughout this paper, the following parameters and  initial condition are set 
$m_i = m = 1$, $\ell_i= \ell = 1$ for all $i$ 
and
$x_i = i-(N+1)/2$, $y_i = 0$ for all $i$
, respectively.
The values of system size and initial momentum 
for each simulation set
will be  defined in the subsequent sections.

\section{Result}
Next, we
 briefly summarize the relation between our numerical simulation and
thermal equilibrium.
Under
 most of the initial conditions 
considered for the present simulations,
the system undergoes chaotic motion. 
Since the model has no conserved quantities other than the total energy,
one may think that  the
states of the system attained 
 in the course of 
a long-duration are
 well approximated by a microcanonical distribution.
In that case, the distribution of the state of each particle in the chain can be
approximated by a canonical distribution at a certain temperature,
by considering the other particles in the chain as a heat bath.
That is, the long-term average of kinetic energy of each particle, $\overline{K_i}(t)$,
is equal to thermal average $\langle K_i \rangle$.
Then, according to the principle of equipartition of energy,
the average kinetic energy of each particle is  the same:
\begin{align}
  \overline{K_i}(t)
&\equiv 
\frac{1}{t}\int_0^t \frac{m_i}{2}\left(\dot x_i(t')^2 +\dot y_i(t')^2\right)dt' \nonumber\\
&\underset{t\rightarrow \infty}{\longrightarrow } \langle K_i \rangle \equiv \frac{1}{Z}\int K_i \exp\left(-\beta H\right)dpdq
\nonumber\\
&= k_B T \, ,  \label{eq:equipartition} 
\end{align}
where $Z\equiv \int  \exp\left(-\beta H\right)dpdq$, $\beta\equiv 1/k_BT$,
$k_B$ is the Boltzmann constant, and $T$ is the temperature.
Since our aim is to investigate the property of relaxation to equipartition,
we define the following quantity in order  to measure how close the system is
to 
equipartition:
\begin{equation}
  \label{eq:delta}
  \Delta(t) \equiv \frac{1}{N}\sum_{i=1}^N \left[\overline{K_i}(t) 
  - \left(\frac{1}{N}\sum_{i'=1}^N\overline{K_{i'}(t)}\right)\right]^2 \, .
\end{equation}
If $\Delta(t)=0$, then $\overline{K_i}(t) =K_0$ for all $i$.
Similar quantities have been used to measure the degree of equipartition
in  studies on a  supercooled liquid~\cite{ergodic-measure-kirkpatric},
self-gravitating systems~\cite{sheet-tkg-1,sheet-tgk-2,sheet-tgk-3},
and proteins~\cite{ergodic-measure-straub}.


\begin{figure}
    \centering
   \includegraphics[width=0.95\columnwidth]{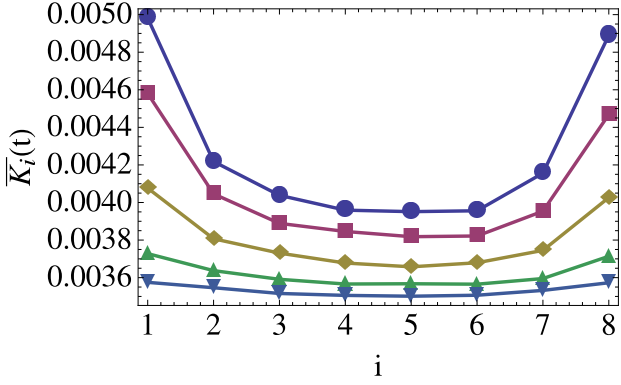}
  \caption{$\overline{K_i}(t)$ vs. $t$ for $k=10^4$. $N=8$.
    Plots are measured  at 10 successive times: $t=10^5$ (blue  circles), $2\times 10^5$ (red squares), 
$4\times 10^5$ (yellow diamonds), $8\times 10^5$ (green triangles),  
and  $1.6\times 10^6$ (blue inverted triangles).
    The time step of integration $\delta t=4\times10^{-4}$. 
The initial conditions are $x_i=i - (N+1)/2$, $y_i=0$, $p^{x}_i=0$,  and $p^{y}_1=0.1$,
$p^{y}_i=-0.1$ $(i\ge 2)$.
}
  \label{fig:ki-vs-t-for-k-1e4}
\end{figure}

Figure \ref{fig:ki-vs-t-for-k-1e4}  shows the time evolution of the profile of 
$\left\{ \overline{K_i}(t), \ i=1,2,\cdots, N=8\right\}$.
It is clearly observed that in the  initial stage of the time evolution,
the average kinetic energy of all particles is not equal; 
rather, 
particles near both the ends of the chain have a  larger average $K_i$.
The profile is similar to that of the rigid link, i.e., the planar 
chain model~\cite{chain-letter-JSTAT-2009}.
Then, as  time progresses,  differences in  average $K_i$ among particles
gradually decrease and tend to zero, and equipartition is achieved.
\begin{figure}
  \centering
 \includegraphics[width=0.95\columnwidth]{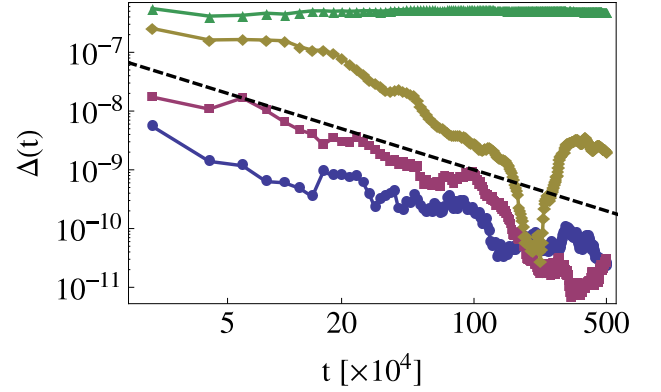}
  \caption{Time evolution of $\Delta(t)$  for $k=1$ (blue circles), 
$10$ (red squares), $10^4$ (yellow diamonds), and $10^5$ (green triangles).
 The other parameters and initial conditions are the same  as
those mentioned in the caption of  Fig.~\ref{fig:ki-vs-t-for-k-1e4}.
The dashed line represents $0.001/t$.
}
  \label{fig:delta-vs-t-k1e4}
\end{figure}
Figure \ref{fig:delta-vs-t-k1e4}  shows the relaxation of $\Delta(t)$ [Eq.~(\ref{eq:delta})] for
the data considered in   Fig.~\ref{fig:ki-vs-t-for-k-1e4}.
We observe that the system relaxes to equilibrium  with the progress of  time.

The physical process of relaxation can be understood by examining the 
kinetic energy in greater detail. 
We rewrite the Hamiltonian as
\begin{eqnarray}
  \label{eq:hamiltonian-vib-rot}
  H &=& K_{vib}\bigl(\VEC{\dot \ell}\bigr) 
  + K_{rot}\bigl(\VEC{\dot\varphi}\bigr) 
  + K_{int}\bigl(\VEC{\dot \ell},\VEC{\dot \varphi}\bigr)  + U(\VEC{r})\\
   K_{vib} &\equiv& 
   \frac{M}{2}\sum_{j,k=1}^{N-1}\mu_{\min(j,k)}^{\le} \mu_{\max(j,k)}^{>} 
   \dot\ell_j \dot\ell_k \cos(\varphi_j - \varphi_k) \, ,\nonumber\\
   K_{rot} &\equiv& 
   \frac{M}{2}\sum_{j,k=1}^{N-1}\mu_{\min(j,k)}^{\le} \mu_{\max(j,k)}^{>} 
   \ell_j \ell_k \dot\varphi_j\dot\varphi_k\cos(\varphi_j - \varphi_k) \, ,\nonumber \\
   K_{int} &\equiv& 
   M\sum_{j,k=1}^{N-1}\mu_{\min(j,k)}^{\le} \mu_{\max(j,k)}^{>} 
   \dot \ell_j \ell_k \dot\varphi_k\sin(\varphi_j - \varphi_k) \, ,\nonumber
\end{eqnarray}
where 
\begin{equation}
M\equiv \sum_{i=1}^N m_i, \ \mu_k\equiv \frac{m_k}{M}, \   \  \mu_n^{\le} \equiv \sum_{k=1}^n \mu_k \ ,  \  \ 
 \mu_n^{>} \equiv \sum_{k=n+1}^N \mu_k \, .
\label{eq:mu-gl}
\end{equation}

\begin{figure}
  \centering
  \includegraphics[width=0.95\columnwidth]{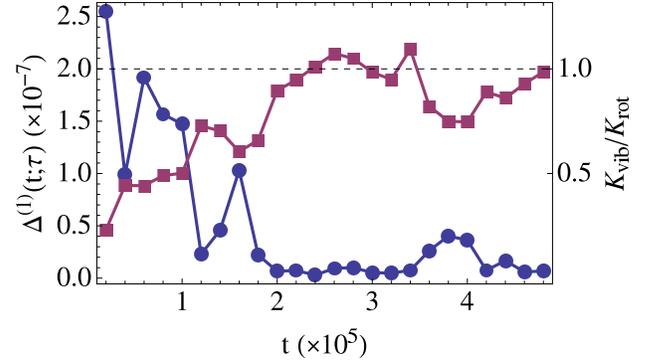}
    \caption{Time evolution of $\Delta^{(1)}(t;\tau)$ (blue circles) and 
$K_{vib}/K_{rot}$ (red squares).
    $N=8$, $K=100$, and $\delta t=10^{-3}$. $\tau=2\times 10^3.$
    The initial conditions are
    $x_i=i-(N+1)/2$, $y_i=0$, $p^{x}_i=0$, $p^{y}_i=0.1$ ($i=1$), and $p^{y}_i=-0.1$ $(2\le i \le 8)$ . }
  \label{fig:delta-kvib-krot-vs-t}
\end{figure}
If equipartition is achieved, $\langle K_{vib}\rangle = \langle K_{rot}\rangle$,
because the model has the same number of springs and angles.
Figure~\ref{fig:delta-kvib-krot-vs-t} shows the temporal evolution of 
$\Delta^{(1)}(t;\tau)$ and $\overline{K_{vib}}(t;\tau)/\overline{K_{rot}}(t;\tau)$ 
for $k=200$.
Here, the time average with two arguments, $\overline f(t;\tau)$,
is defined as 
\begin{equation}
  \overline f(t;\tau)\equiv \frac{1}{\tau}\int_t^{t+\tau}f(t')dt' \,  ,
\end{equation}
and 
\begin{equation}
  \Delta^{(1)}(t;\tau) \equiv \frac{1}{N}\sum_{i=1}^N \left[\overline{K_i}(t;\tau)
  - \left(\frac{1}{N}\sum_{i'=1}^N\overline{K_{i'}(t;\tau)}\right)\right]^2 \, .
\end{equation}
We find that the system relaxed to equipartition 
on a timescale similar to that on which
the rotational energy $K_{rot}$ transformed into  vibrational energy $K_{vib}$.


As mentioned earlier,
the aim of this study is to examine the property of relaxation to 
equipartition of energy for the model expressed in Eq.~(\ref{eq:spring-chain}).
First, for each sample orbit starting from different initial condition, 
we define the relaxation time  $t_{relax}^{(sample)}$ 
as the time required for $\Delta(t)$ to decay below a
critical value $\Delta_0$. 
We define the average relaxation time over $N_{sample}$ orbits  as
\begin{equation}
  \label{eq:t-relax}
 t_{relax}\equiv \frac{1}{N_{sample}}\sum_{sample}  t_{relax}^{(sample)}.
\end{equation}
This  time is a measure of the relaxation time to equipartition.

\begin{figure}
  \centering
  \includegraphics[width=0.95\columnwidth]{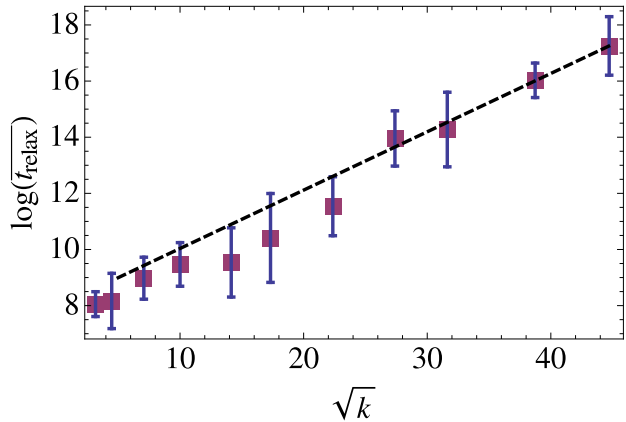}
  \caption{$k$ dependence of average relaxation time $t_{relax}$.
At each $k$ value, 15 samples are taken.
Threshold value $\Delta_0 = 10^{-7}$.
  The time step $\delta t$ of integration  is scaled as $\delta t=0.125*\sqrt{10}/\sqrt{k}$.
  The error bars represent standard deviations of $t_{relax}^{(sample)}$ for 15 samples.
  The dashed line shows 
$t_{relax} = 5.52\times 10^4\exp(0.415 \sqrt{k})$
obtained by fitting.}
  \label{fig:log-trelax-vs-sqrt-k}
\end{figure}
Figure~\ref{fig:log-trelax-vs-sqrt-k} shows 
the plot of 
the dependence 
of $t_{relax}$ on the spring constant $k$. 
We observe that as the stiffness of the spring increases,
the relaxation time  increases rapidly. That is,  systems with hard springs
or a steep potential show  rigid-like behavior
of energetic particles near the chainends 
for a very long time,
as shown in Fig.~\ref{fig:ki-vs-t-for-k-1e4}, before relaxing to equipartition.

Here, we mention  
a technical detail about the numerical integration used for
obtaining the plot in 
Fig.~\ref{fig:log-trelax-vs-sqrt-k}.
With increasing spring constant $k$, the period of bond-stretching vibration
decreases
in proportion to $1/\sqrt{k}$. 
Therefore, for large values of $k$, 
the magnitude of the time step of numerical integration
should be reduced.
We confirmed that $t_{relax}$ converges at  $\delta t=0.125$
for $k=10$.
Thus, we used $\delta t=0.125*\sqrt{10}/\sqrt{k}$ for each $k$.

On changing the 
initial conditions, the relaxation time $  t_{relax}^{(sample)}$ changes, 
and we obtain a distribution of  $t_{relax}^{(sample)}$,  denoted as $P(t_{relax})$.
\begin{figure}
  \centering
  \includegraphics[width=0.95\columnwidth]{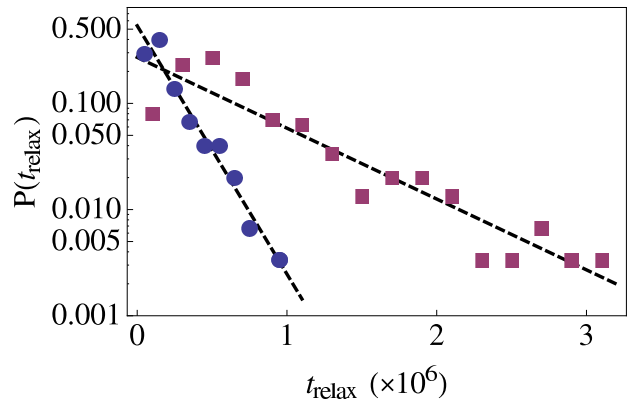}
  \caption{Distribution of $t_{relax}$ for $k=100$ (blue circles) and $k=200$ (red squares).
For each value of $k$, 
 the numbers of samples are 200.
Both distributions are shown on a  semi-log scale. 
The dashed lines  show
$P(t_{relax})= c\exp(-\alpha t_{relax})$. 
For $k=100$, $c=0.544$ and $\alpha=5.39\times 10^{-6}$.
For $k=200$, $c=0.271$ and $\alpha=1.54\times 10^{-6}$.
}
  \label{fig:hist-trelax}
\end{figure}
Figure~\ref{fig:hist-trelax} shows the distribution of $t_{relax}$ for
$k=20$ and $k=100$.
For both cases, the histograms show exponential decay expressed as
\begin{equation}
  \label{eq:hist-exp}
  P(t_{relax}) \propto \exp(-\alpha t_{relax}) \,,
\end{equation}
which suggests the existence of a characteristic timescale for the relaxation.
 
Next,  we analyze the results of the abovementioned calculation
using the concept  of the Boltzmann-Jeans theory (also known as Boltzmann-Jeans conjecture)~\cite{benettin-jeans-2,benettin-jeans-ptp-94,benettin-jeans-nonlin-96,benettin-galgani-giorgilli-pla-1987,jeans-numeric-91,jeans-93}.
The essence of this theory is roughly described as follows. 
( For a detailed description
of the theory please refer to \cite{benettin-jeans-2}. )
Suppose we have a system described by a Hamiltonian, which has
two subsystems $H_f$ and $H_s$, their typical time scale being
$\tau_f$ and $\tau_s$, respectively. Here, subscripts $f$ and $s$ 
denote ``fast'' and ``slow,'' respectively. Let us call 
$H_f$ and $H_s$ as a ``fast subsystem'' and ``slow subsystem,'' respectively.
If the timescales of the fast and slow subsystems differ greatly, i.e.,
$\tau_s/\tau_f \gg 1$, then the timescale for the occurrence of energy exchange
between these two subsystems is on the order of
\begin{equation}
  t_{exch} \gtrsim \exp(c\frac{\tau_s}{\tau_f}) \, . 
\end{equation}
That is, 
energy exchange occurs after a long time.
In the case of the spring-chain system ~[Eq. (\ref{eq:spring-chain})] with a 
large spring constant $k$, the fast and slow subsystems correspond to 
bond vibration and relative rotation, respectively.
Since the typical timescale of bond vibration is on the order of 
$2\pi\sqrt{m/k}$ and that of rotation is assumed to be constant,
we have
\begin{equation}
  t_{exch} \sim \exp(c\cdot 1/\sqrt{\frac{m}{k}} ) =\exp(c'\sqrt{k}) \, .
  \label{eq:timescale}
\end{equation}
Since the relaxation to equipartition occurs by energy transfer
from rotation to vibration (as we observed before),
we can consider that  $t_{exch}$ mentioned above is essentially the same as
the relaxation time $t_{relax}$:
\begin{equation}
  t_{relax} \sim \exp(c'\sqrt{k}) \, . 
  \label{eq:timescale-relax}
\end{equation}

Now, we examine whether $t_{relax}$ obtained by the simulation obeys 
Eq.~(\ref{eq:timescale-relax}).
The result is already shown in  Fig.~\ref{fig:log-trelax-vs-sqrt-k}.
$\mbox{Log} (t_{relax})$ is proportional to $\sqrt{k}$; therefore the interpretation 
by the Boltzmann-Jeans theory is appropriate.

This  theory can also be used for interpreting the histogram of
$t_{relax}^{(sample)}$. Since $t_{relax}$ is defined from the average of a number of
samples, we have
\begin{equation}
  t_{relax} = \int_0^\infty t_{relax}' P(t_{relax}')dt_{relax}' \,,
\end{equation}
where $P(t)$ is the distribution of $t_{relax}^{(sample)}$.
If we adopt the exponential form for the distribution $P$~[Eq.~(\ref{eq:hist-exp})],
then
\begin{equation}
  t_{relax} = \frac{1}{\alpha} \, ,
  \label{eq:t-relax-alpha}
\end{equation}
where $\alpha$ is the coefficient that appears in Eq.~(\ref{eq:hist-exp}).

Combining Eqs.~(\ref{eq:timescale-relax}) and (\ref{eq:t-relax-alpha}),
the relation between the coefficient $\alpha$  and the spring constant
should be
\begin{equation}
  \alpha\propto \exp(-c'\sqrt{k}) \, .
  \label{eq:alpha-vs-k}
\end{equation}

\begin{figure}
  \centering
  \includegraphics[width=0.95\columnwidth]{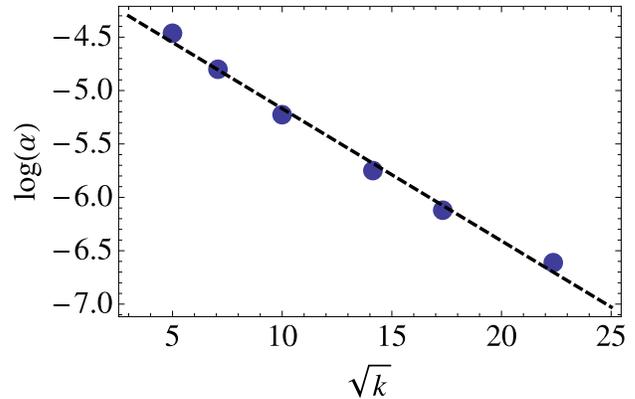}
  \caption{Plot of $\alpha$ vs $k$,   obtained from 300 samples.
 The dashed line shows $\log \alpha = c_1 - c_2 \sqrt{k}$, where
$c_1=-3.93$ and $c_2=0.124$.}
  \label{fig:alpha-vs-k}
\end{figure}

A comparison between the estimation~[Eq.~(\ref{eq:alpha-vs-k})] and data is 
shown in Figure \ref{fig:alpha-vs-k},
from which we find that they are in 
good agreement.
Thus, 
the fact that relaxation to equilibrium takes quite a long time to occur
can be interpreted as the outcome of the Boltzmann-Jeans theory.

\section{Summary and Discussion}
In this paper, we have numerically shown the occurrence of  energetic motion of end particles
 for a 1D chain of point masses connected by hard springs.
The timescale at  which the energetic motion is observed depends on the 
spring constant, and this timescale 
lengthens with increasing
the spring constant.
Relaxation to equilibrium occurs as rotational kinetic energy
is converted into vibrational kinetic energy.

The timescale of relaxation is estimated using the Boltzmann-Jeans theory, 
which describes the energy exchange rate in a system in which
fast  and slow motions coexist. In the case of our model,
fast motion corresponds to  bond vibration by a hard spring and slow motion 
corresponds to rotation and deformation of the chain.
The result of our numerics is that the timescale is estimated as an exponential of
the square root of the spring constant, which coincides well with simulation data.

The energetic motion of end particles is also observed in other systems, 
e.g., a planar chain system~\cite{chain-letter-JSTAT-2009}
and multiple pendulum~\cite{yanagita-gakkai-1,yanagita-gakkai-2,yanagita-gakkai-3}.
These systems consist of masses connected by rigid links,
and therefore,  they are constrained systems.
If we consider a rigid link as a limiting case of a hard spring
when the spring constant $k$ is large ($k\rightarrow\infty$),
then the spring-chain system examined in this study becomes
a planar chain system when $k\rightarrow \infty$.
Therefore, it is natural that the spatial energy distribution of the spring-chain system
considered in this study resembles those of  systems with rigid links.


However, the equilibrium behaviors of spring-chain system and rigid-link chain  system
are quite different. In the spring-chain  system, equipartition achieved is of 
 the usual form; that is, 
the average kinetic energy of each particle is the same, regardless of the 
magnitude of  $k$. Then, 
chain-end particles in thermal equilibrium
should not exhibit the energetic behavior.
In contrast, in the rigid-link-chain system, 
chain-end particles behave energetically even in thermal
equilibrium~\cite{chain-letter-JSTAT-2009}.
The energetic behavior of end particles in the spring-chain system is a transient 
behavior before the system relaxes to thermal equilibrium.
What is important here is that the relaxation time increases
with increasing 
the  spring constant $k$
and that   the relaxation time eventually diverges at $k \rightarrow \infty$.
Therefore,  a large $k$ value provides a good opportunity 
to observe energetic behavior in spring-chain systems.

The relaxation time to equipartition changes if we change the initial conditions.
We found that the  the relaxation time is distributed according to  the exponential form
for large relaxation time. 
This implies that  relaxation occurs almost randomly. 
That is, the system moves on the energy surface in a random way
and happens to encounter at which  the system can divert 
toward the equipartition state.
This situation is in contrast to  the process of  slow dynamics often observed
in many Hamiltonian systems,
where the relaxation time is often distributed according to the power law.


Slow relaxation is often observed in Hamiltonian systems,
and in most cases, it is accompanied by strong temporal correlation caused by
sticky or stagnant motion around KAM tori (regular orbits) and their remnants
and $1/f$-type fluctuations. Such slow relaxation is observed in 
area-preserving mappings and some other high-dimensional 
systems~\cite{Karney,Chirikov-Shepel,Kohyama,Geizel,MarkovTreeModel,meiss-islands,Aizawa-prog,Aizawa_selfsimilar,yyama-transport}.
Such slow relaxation is explained by a hierarchical structure generally found in 
nearly integrable Hamiltonian systems. In this sense, it is quite common to find
slow relaxation in Hamiltonian systems, if the Hamiltonian is nearly integrable.

However, the spring-chain model considered in this study is not nearly integrable.
In addition,  the systems which show slow relaxation in real world 
are not always nearly integrable.
In contrast, 
the Boltzmann-Jeans theory is still applicable to  Hamiltonian systems
that are not nearly integrable and is able to explain the slow relaxation.
Similar to the research by Shudo et al.~\cite{shudo-saito-bj}, 
the result of our study also implies that the Boltzmann-Jeans theory
can describe slow relaxation in more general systems 
other than the nearly integrable ones.

We used the Boltzmann-Jeans theory~\cite{benettin-jeans-2,benettin-jeans-ptp-94,benettin-jeans-nonlin-96,benettin-galgani-giorgilli-pla-1987,jeans-numeric-91,jeans-93}
to estimate the relaxation time.
Although the concept of the theory dates back to the 19th century~\cite{boltzmann-1895,jeans-1903,jeans-1905}, 
its importance is not very familiar and  not  many 
examples of the application of 
this theory have been demonstrated~\cite{shudo-saito-bj,nakagawa-kaneko-jpsj-2000,nakagawa-kaneko-jpsj-2000-2,nakagawa-kaneko-pre-2001,morita-kaneko-epl-2004,morita-kaneko-prl-2005}.
Because  the results of this study show that there is 
 good agreement between numerical data and theoretical estimation,
this study can be a good example of  the applicability of the  Boltzmann-Jeans theory.

In this study, we showed that chain-end particles behave energetically
even for systems with a finite but large spring constant.
This implies that similar behavior can be observed for 
natural chain-type systems that are not made of
rigid links but whose intrachain potential 
between elements is very steep. 
The results of this study are expected to have many useful applications to 
polymers, proteins and some artificial objects.

\begin{acknowledgments}
This study was partially supported by a Grant-in-Aid for Scientific Research (C)
(20540371) from the Japan Society for the Promotion of Science (JSPS).
\end{acknowledgments}



\end{document}